\documentstyle[12pt,epsfig]{article}
\begin{document}
\begin{flushright}
PPREPRINT IHEP-98-87\\
hep-ph/9812348
\end{flushright}

\begin{center}

{\Large \bf The NLO DGLAP extraction of $\alpha_s$ and higher twist 
terms from CCFR $xF_3$ and $F_2$ structure functions data for $\nu N$ DIS}

\vspace{0.1in}

{\bf S.I. Alekhin$^{a}$ and A.L. Kataev$^{b}$\footnote{On leave of absence 
at December 1998 from Institute for Nuclear Research of the Academy of 
Sciences of Russia, 117312, Moscow, Russia}}

\vspace{0.1in}
{\baselineskip=14pt
$^{a}$Institute  
for High Energy Physics, 142281 Protvino, Russia}

\vspace{0.1in}
{\baselineskip=14pt
$^{b}$ Theory Division, CERN, CH-1211 Geneva 23, Switzerland}

\begin{abstract}
We performed the detailed NLO analysis of the combined CCFR $xF_3$ 
and $F_2$ structure functions data and extracted the value of 
$\alpha_s$, parameters of distributions and higher-twist (HT) terms 
using the direct solution of the DGLAP equation. The value 
of $\alpha_s(M_Z)=0.1222\pm 0.0048 (exp)\pm 0.0040(theor)$ was obtained.
The result has larger central value and errors, than the original 
result of the CCFR collaboration, in view of the incorporation into the 
fits of the HT terms as the free model independent parameters. 
The $x$-shapes of the HT
contributions 
to  $xF_3$ and $F_2$ are in agreement with the results of other 
model-independent extractions and are in  
qualitative agreement with the predictions of the infrared renormalon 
model. We also argue that the low $x$ CCFR data might have the defects,
since their inclusion into the fits led to the following low
$x$-behaviour of the gluon distribution $xG(x,9~GeV^2) \sim x^{0.092\pm
0.0073}$, in contradiction with the results of its extraction from low $x$ 
HERA data. 
\end{abstract}
\end{center}
\newpage
{\bf 1.} The study of the possibility to separate 
power suppressed terms (namely higher-twist (HT) effects) from 
the perturbation theory 
logarithmic corrections  in  the analysis 
of scaling   violation of the 
deep-inelastic scattering (DIS) processes has a rather long 
history (see e.g. Refs. \cite{BB1,AB,Isaev} and Ref.\cite{Buras}
for the review). In the recent years the interest to this 
problem was renewed, mainly due to the consideration  of the 
possibility to model HT terms in different processes using the
infrared-renormalon (IRR)  technique   
(see e.g. Refs.\cite{BB}-\cite{BBM},\cite{B} 
and especially Ref.\cite{Beneke} for the review).

On the other hand the experimentalists are improving 
the precision of their data and are achieving, sometimes, percent level 
of accuracy. For example the data on $xF_3$ and $F_2$  
from the most precise $\nu N$ DIS experiment, 
performed at Tevatron by CCFR collaboration, recently appeared
\cite{CCFR,SELIG}. The CCFR data on $xF_3$ were 
analysed in Ref.\cite{KKPS} in the LO and and with inclusion of the NLO 
and approximate next-to-next-to-leading order (NNLO) corrections. 
For the latest,the   
NNLO QCD 
corrections  to the coefficient function \cite{VZ}
were taken into account. 
The  NNLO corrections to the  
anomalous dimensions of a limited set of even non-singlet moments 
\cite{LRV} were also taken into account. 
The NNLO corrections to the anomalous dimensions of odd moments, 
which are not still explicitly calculated, were obtained using
smooth interpolation procedure
proposed in Ref.\cite{PKK} and improved in Ref.\cite{KKPS1}.
The aim of the work of Ref.\cite{KKPS} was to 
make an attempt of the first NNLO  determination of  
$\alpha_s(M_Z)$ from  DIS and to extract the 
HT terms from the data 
on $xF_3$ within the framework of the IRR-model \cite{DW} and
also by the model-independent way, similarly to the analysis 
of the combined SLAC/BCDMS data \cite{VM}, performed in the NLO approximation.
Theoretical uncertainties of this analysis were further estimated 
in Refs.\cite{KKPS2,KPS} at the N$^3$LO 
using the method of Pad\'e approximants. 
It was found in Refs. \cite{KKPS,KKPS2,KPS} that the inclusion of
the NNLO corrections leads to the decrease
of the HT contribution value, so that at the  NNLO
its $x$-shape lies closer to zero.

In these analyses  only statistical errors of data were taken into account.
However, the systematic errors of the CCFR experiment are not small 
\cite{SELIG} and may even dominate in the determination of some
parameters. In this paper we  filled in this gap and performed the 
NLO analysis of the CCFR data with the help of QCD DGLAP evolution code,
developed in Ref.\cite{ALEK}. (Remind, that the analyses 
of Refs.\cite{KKPS,KKPS2,KPS} 
were performed with the help of the Jacobi 
polynomial variant \cite{Jacobi1,Jacobi2,Jacobi3} of the DGLAP equation
\cite{DGLAP}). In addition, we  included in our 
analysis the CCFR  data on the singlet structure function $F_2$.
It should be stressed that  
the code \cite{ALEK} was tested using the procedure 
proposed in Ref.\cite{BENCH} and demonstrated the accuracy at the level of 
$O(0.1\%)$ in the kinematic region covered by the analysed data.
It was already applied for the nonsinglet DGLAP analysis of 
the combined SLAC/BCDMS SLAC data on $F_2$ \cite{Aln}. 

{\bf 2.} Our fits were made in the NLO approximation within the 
$\overline{MS}$ factorization and renormalization schemes. The $Q^2$
dependence of the strong coupling constant $\alpha_s$ 
was defined from the following equation
\begin{equation}
\frac{1}{\alpha_s(Q)}-\frac{1}{\alpha_s(M_Z)}=
\frac{\beta_0}{2\pi}\ln\biggl(\frac{Q}{M_Z}\biggr)+
\beta\ln\biggl[\frac{\beta+1/\alpha_s(Q)}{\beta+1/\alpha_s(M_Z)}\biggr],
\label{ALPHA}
\end{equation}
where
$\beta=\frac{4\pi\beta_1}{\beta_0}$
and $\beta_0$ and $\beta_1$ are  the coefficients of the QCD
$\beta$-function, 
defined as 
\begin{equation}
\beta(\alpha_s)=\frac{1}{4\pi}\mu\frac{\partial\alpha_s}{\partial\mu}=
-2\sum_{i\geq 0} \beta_i (\frac{\alpha_s}{4\pi})^{i+2}
\end{equation}
where $\beta_0=11-(2/3)n_f$  
and $\beta_1=102-(38/3)n_f$.
Note, that the explicit solution of Eq.(1) can be expressed 
through the Lambert function \cite{GGK}. However, we did not use 
in our work this explicit representation and solved Eq.(1) 
numerically. 
The effective number of flavours $n_f$
was chosen to be $n_f=4$ for $Q^2$ less than the definite scale $M_5^2$ 
and increased to $n_f=5$ at larger values of $Q^2$
keeping the continuity of $\alpha_s$ \cite{BW}.
The value of the effective matching scale $M_5$ was varied from 
$M_5=m_b$ to $M_5=6.5m_b$. 
The last  choice
was advocated in Ref.\cite{BN} on the basis of the DIS sum rules consideration. 
The dependence of the results of the 
fits   on the choice of the matching point 
gives one of the sources of theoretical uncertainties inherent 
to our analysis.

The leading twist term $xF_3^{LT}(x,Q)$ 
was obtained 
by direct integration of the DGLAP 
equation \cite{DGLAP}
\begin{equation}
\frac{dxq^{NS}}{d\ln Q}=\frac{\alpha_s(Q)}{\pi}\int^1_x
dzP^{NS}_{qq}(z)\frac{x}{z}q^{NS}(x/z,Q),
\end{equation}
where $P^{NS}_{qq}(x)$ 
denotes the NLO splitting function, taken from Ref.\cite{MS}.
The function $xF_3$ is determined by the  
 subsequent convolution with the NLO coefficient function $C_{3,q}(x)$
:
\begin{equation}  
xF_3^{LT}(x,Q)=\int_x^1dz
C_{3,q}(z)\frac{x}{z}q^{NS}(x/z,Q).
\end{equation}  
The boundary condition
at the reference scale $Q_0^2=5~GeV^2$
was chosen in the form 
analogous to the ones, used in Refs.\cite{SELIG,KKPS}
\begin{equation}  
xq^{NS}(x,Q_0)=\eta_{NS}x^{b_{NS}}(1-x)^{c_{NS}}(1+\gamma x)\frac{3}{A_{NS}},
\end{equation}
where 
\begin{equation}  
A_{NS}=\int_0^1x^{b_{NS}-1}(1-x)^{c_{NS}}(1+\gamma x)dx,
\end{equation}  
and  $\eta_{NS}$  
is the measure of the deviation of the Gross-Llewellyn Smith 
integral \cite{GLS} from  its quark-parton value 3.
The expression for the $xF_3$, which includes the 
HT contribution, looks as follows:
\begin{equation}  
xF_3^{HT}(x,Q)=xF_3^{LT,TMC}(x,Q)+\frac{H_3(x)}{Q^2}, 
\end{equation}  
where $F_3^{LT,TMC}(x,Q)$ 
is $F_3^{LT}(x,Q)$
with the target mass correction \cite{TMC} applied. 

At the first stage of this work, in order to perform the
cross-checks of the code against the results of Refs.\cite{SELIG,KKPS},
we fitted the data on $xF_3$ in the kinematical region 
$Q^2>5~ GeV^2$, $W^2>10~GeV^2$, $x<0.7$ (the number of data points is NDP=86).
We made three fits with various
ways  of taking into account the HT effects.
The first fit with 
no HT, i.e. $H_3(x)=0$, was done 
to compare our results with the ones of  Table 1 of Ref.\cite {KKPS}, 
obtained using  
different method \cite{Jacobi1,Jacobi2,Jacobi3} and different computer code.
The second fit with the HT chosen 
as one-half of the IRR model predictions \cite{DW}, i.e.
\begin{equation}  
H_3(x)=A'_2\int_x^1dz
C_2^{IRR}(z)\frac{x}{z}F_3^{LT}(x/z,Q),
\end{equation}  
where
\begin{equation}  
C_2^{IRR}(z)=-\frac{4}{(1-z)_+}+2(2+z+2z^2)-5\delta(1-z)-\delta'(1-z)
\end{equation}  
and $A'_2=-0.1~GeV^2$, advocated for the first time in Ref.\cite{Stein}.
The aim of this fit 
was to compare its outcomes with the results of Table 7.9 of Ref.\cite{SELIG}, 
where the computer code, written by Duke and Owens \cite{DO} was used.
In the  third fit we used  the model independent HT-expression, i.e.
$H_3(x)$ parametrized
at $x=0.,0.2,$ $0.4,0.6,0.8$ with linear interpolation 
between these points. It was performed to compare our results with 
the ones presented in Table 3 of Ref.
\cite{KKPS}.

All results of these our fits are presented in Table 1. We observed a good
agreement 
of our results on $\alpha_s$ with the both referenced papers.
However, in the case of the values of $x$-shape parameters 
we found the certain discrepancy with the results of Ref.\cite{KKPS}.
For example, 
the value of $\gamma$, as presented in column I of Table 1, is
$\gamma=0.26\pm0.30$, meanwhile the analogous parameter in Ref.\cite{KKPS}
is $\gamma=1.96\pm0.36$. At the same time 
our $x$-shape parameters are in agreement with the ones, extracted 
in Ref.\cite {SELIG} 
within errors.
In addition, we  made the fit, releasing parameter $A_2^{'}$ and obtained 
the value of $A_2^{'}=-0.12\pm0.05$ in agreement with the results 
of Ref.\cite{KKPS}.

\begin{table}
\caption{The results of the fits to data on $xF_3$ with statistical errors only
I) without HT-terms,
II) with HT accounted as one-half of the IRR model predictions,
III) with model independent HT-contributions. $H_3^{(0),(2),(4),(6),(8)}$ 
are the values of $H_3(x)$ at $x=0.,0.2,0.4,0.6,0.8$.}

\begin{tabular}{cccc} 
                & I                 & II                & III               \\
$\chi^2/NDP$    & 88.5/86           & 81.9/86           & 70.3/86           \\
$b$             & $0.789\pm0.024$   & $0.786\pm0.024$   & $0.805\pm0.067$   \\
$c$             & $4.02\pm0.11$     & $4.00\pm0.11$     & $4.24\pm0.21$     \\
$\gamma$        & $0.29\pm0.30$     & $0.26\pm0.30$     & $0.61\pm0.71$     \\
$\eta_{NS}$     & $0.927\pm0.014$   & $0.949\pm0.014$   & $0.927\pm0.030$   \\
$\alpha_s(M_Z)$ & $0.1193\pm0.0025$ & $0.1219\pm0.0024$ & $0.1216\pm0.0066$ \\
$H_3^{(0)}$     &            --     &  --               & $0.18\pm0.19$       \\
$H_3^{(2)}$     &            --     &  --               & $-0.26\pm0.12$    \\
$H_3^{(4)}$     &            --     &  --               & $-0.21\pm0.31$      \\
$H_3^{(6)}$     &            --     &  --               & $0.11\pm0.26$     \\
$H_3^{(8)}$     &            --     &  --               & $0.90\pm0.47$     \\
\end{tabular}
\end{table}

{\bf 3.} The next step of our analysis was to take into account 
the point-to-point correlations of the data due to 
systematic errors which,
as we mentioned above, can be crucial for the estimation of full
experimental errors of the 
parameters (see in particular \cite{Aln}, where the value
$\alpha_s(M_Z)=0.1180\pm 0.0017~(stat+syst)$ was obtained as the 
result of the combined fits of the SLAC/BCDMS data with HT included). 
The systematic errors were taken into account
analogously to the earlier works \cite{ALEK,Aln}.
The total number of the independent 
systematic errors sources for the analysed data is 18 and all of them  
were convoluted
into the general correlation matrix, which was used for the construction of
the minimized $\chi^2$. The results of the fits to $xF_3$ data with 
the model independent HT and with the 
systematic errors taken into account are presented in the first 
column of Table 2.
One can see that the  account of 
systematic errors leads to the significant increase of the 
experimental uncertainties  of the extracted HT contributions (compare
the first column of Table 2 with the
third column of Table 1).  
In addition, the central values of the HT parameters moved.
However, even in this case there is definite  
agreement with the results of HT-behaviour of Ref.\cite{KKPS},
obtained  in NLO. Moreover, these results do not contradict to the 
IRR-model prediction of Ref.\cite{DW}, since 
releasing $A_2^{'}$ we obtained $A_2^{'}=-0.10 \pm 0.09$.

\begin{table}
\caption{The results of the fits with the account of systematic errors 
and model independent HT-effects. $H_{2,3}^{(0),(2),(4),(6),(8)}$ 
are the values of $H_2(x)$ and $H_3(x)$ 
$x=0.,0.2,0.4,0.6,0.8$. 
I) $xF_3$ with the cut $Q^2>5~GeV^2$, $Q_0^2=5~GeV^2$
II) $xF_3\&F_2$ with the cut $Q^2>5~GeV^2$, $Q_0^2=9~GeV^2$
III) $xF_3\&F_2$ with the cut $Q^2>1~GeV^2$, $Q_0^2=9~GeV^2$.}
\begin{tabular}{cccc} 
                & I                 & II                & III               \\
$\chi^2/NDP$    & 55.7/86           & 154.9/172         & 204.2/220         \\
$b_{NS}$        & $0.797\pm0.076$   & $0.800\pm0.016$   & $0.782\pm0.014$   \\
$c_{NS}$        & $4.24\pm0.21$     & $4.060\pm0.068$   & $4.131\pm0.056$   \\
$\gamma$        & $0.75\pm0.79$     & $0.$              & $0.$              \\
$\eta_{NS}$     & $0.945\pm0.043$   & $0.922\pm0.027$   & $0.920\pm0.025$   \\
$\alpha_s(M_Z)$ & $0.1269\pm0.0065$ & $0.1248\pm0.0048$ & $0.1131\pm0.0045$ \\
$\eta_S$        & --                & $0.1785\pm0.0077$ & $0.1796\pm0.0065$ \\
$b_S$           & --                & $0.$              & $-0.034\pm0.023$  \\
$c_S$           & --                & $8.37\pm0.21$     & $8.00\pm0.29$     \\
$b_G$           & --                & $0.$              & $0.092\pm0.073$   \\
$c_G$           & --                & $7.5\pm2.6$       & $11.50\pm0.90$    \\
$\eta_G$        & --                & $0.69\pm0.35$     & $1.08\pm0.19$     \\
$H_2^{(0)}$     & --                & $-0.23\pm0.56$    & $0.09\pm0.11$     \\
$H_2^{(2)}$     & --                & $-0.28\pm0.18$    & $-0.239\pm0.094$  \\
$H_2^{(4)}$     & --                & $-0.14\pm0.18$    & $0.17\pm0.13$     \\
$H_2^{(6)}$     & --                & $-0.03\pm0.13$    & $0.204\pm0.097$   \\
$H_2^{(8)}$     & --                & $0.21\pm0.18$     & $0.14\pm0.18$     \\
$H_3^{(0)}$     & $0.28\pm0.21$     & $0.34\pm0.11$     & $0.115\pm0.031$   \\
$H_3^{(2)}$     & $-0.22\pm0.19$    & $-0.24\pm0.16$    & $-0.16\pm0.16$    \\
$H_3^{(4)}$     & $-0.42\pm0.35$    & $-0.22\pm0.22$    & $0.28\pm0.19$     \\
$H_3^{(6)}$     & $-0.09\pm0.28$    & $-0.05\pm0.17$    & $0.19\pm0.15$     \\
$H_3^{(8)}$     & $1.21\pm0.50$     & $0.89\pm0.44$     & $0.88\pm0.44$     \\
\end{tabular}
\end{table}

Trying to minimize the errors of the parameters  we added to the
analysis 
the CCFR data for the structure function $F_2$.
To perform the QCD evolution of $F_2$ one is to involve 
into the analysis the singlet  and
gluon 
distributions:
\begin{equation}  
F_2^{LT}(x,Q)=\int_x^1dz
\Bigl[C_{2,q}(z)\frac{x}{z}\bigl(q^{NS}(x/z,Q)+q^{PS}(x/z,Q)\bigr)
+C_{2,G}(z)\frac{x}{z}G(x/z,Q)\Bigr],
\end{equation}  
The distributions $q^{PS}(x,Q)$ and $G(x,Q)$
were obtained by integrating the system
\begin{equation}
\frac{dxq^{PS}}{d\ln Q}=\frac{\alpha_s(Q)}{\pi}\int^1_x
dz\Bigl[P_{qq}^{PS}(z)\frac{x}{z}q^{PS}(x/z,Q)
+P_{qG}(z)\frac{x}{z}G(x/z,Q)\Bigr]
\end{equation}
\begin{equation}
\frac{dxG}{d\ln Q}=\frac{\alpha_s(Q)}{\pi}\int^1_x
dz\Bigl[P_{Gq}(z)\frac{x}{z}q^{PS}(x/z,Q)
+P_{GG}(z)\frac{x}{z}G(x/z,Q)\Bigr]
\end{equation}
with the boundary conditions
\begin{equation}  
xq^{PS}(x,Q_0)=\eta_{S}x^{b_S}(1-x)^{c_S}/A_{S},
\end{equation}  
\begin{equation}  
xG(x,Q_0)=\eta_{G}x^{b_G}(1-x)^{c_G}/A_{G},
\end{equation}  
where
\begin{equation}  
A_{S}=\int_0^1x^{b_{S}}(1-x)^{c_{S}}dx,
\end{equation}  
\begin{equation}  
A_{G}=\frac{1-<xQ(x)>}{\int_0^1x^{b_{G}}(1-x)^{c_{G}}dx}.
\end{equation}  
and $<xQ(x)>$ is the total momentum carried by quarks.
 
\begin{figure}[t]
\centerline{\epsfig{file=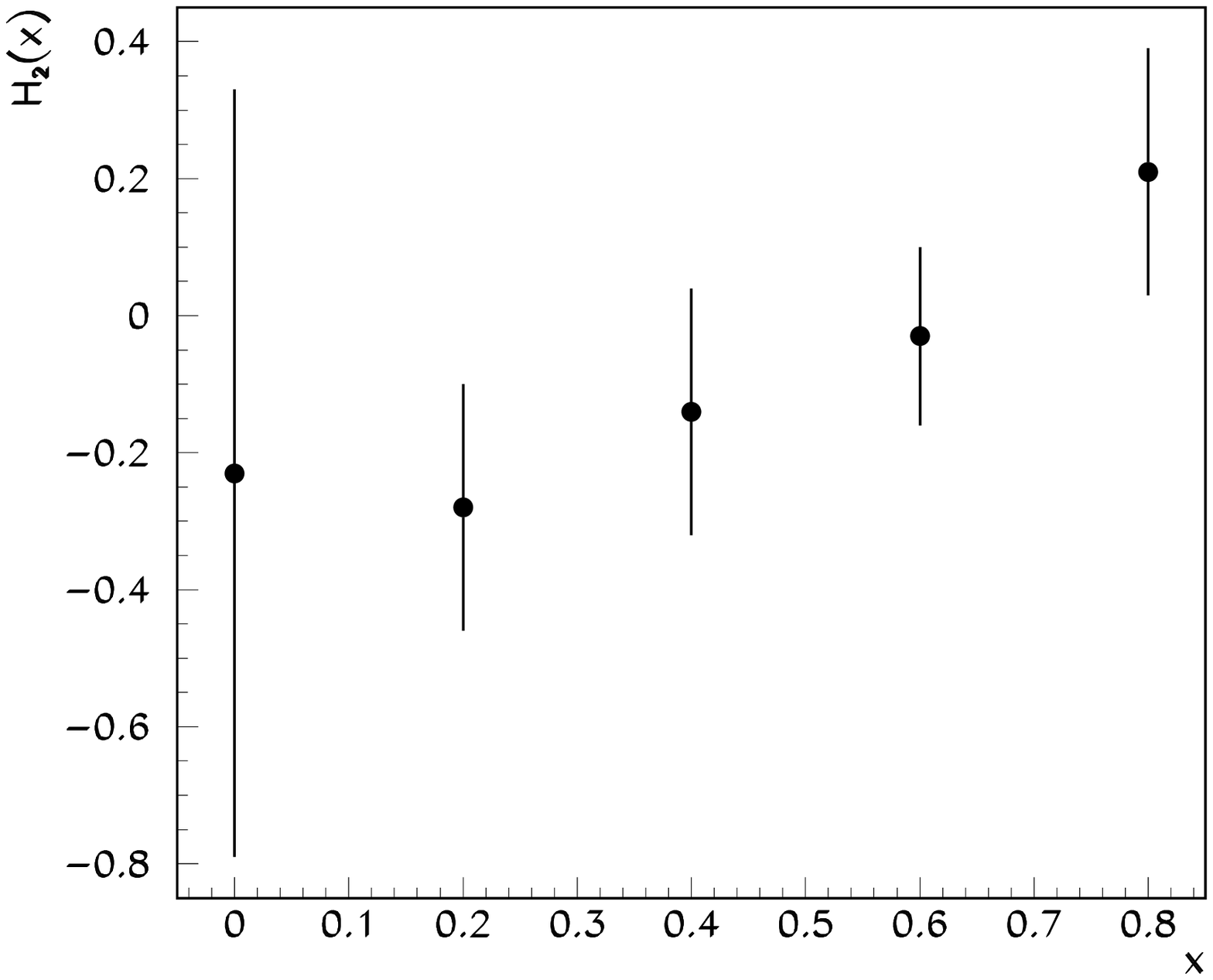,width=8cm}}
\caption{The high-twist contribution to $F_2$.}
\end{figure}
\begin{figure}[t]
\centerline{\epsfig{file=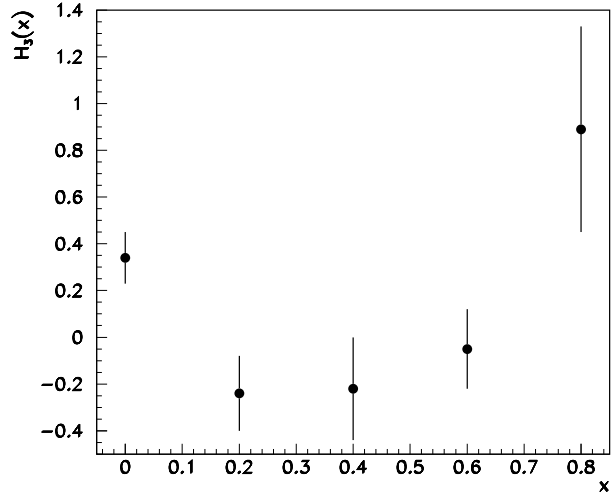,width=8cm}}
\caption{The high-twist contribution to $xF_3$.}
\end{figure}

In order to provide the straightforward way for the comparison of our
results with the analysis of Ref.\cite{ALEK}, the
initial reference scale $Q_0^2$=9 $GeV^2$ was chosen. 
In addition to the point-to-point correlation of the data due to
systematic errors, the statistical correlations between $F_2$ and $xF_3$
were also taken into account.
Performing the trial fits we convinced that 
adding the factor $(1+\gamma x)$ to the reference expressions for the
the gluon and singlet distributions do not improve the quality of the fit.
Also we fixed parameters $\gamma_{NS}, b_S$ and $b_G$ 
at zero because this increased 
the value of $\chi^2$ by few units only 
while $\chi^2/NDP$ remained less than unity. The HT contribution to $F_2$
was accounted analogously to $xF_3$ as:
\begin{displaymath}  
F_2^{HT}(x,Q)=F_2^{LT,TMC}(x,Q)+\frac{H_2(x)}{Q^2}
\end{displaymath}  
where  $H_2(x)$ was parametrized in the model independent form.
The results of the fits of the parameters $H_2(x)$ and $H_3(x)$  are
presented in the second column of Table 2 and are depicted in Fig.1 and
Fig.2.  One can note, that, comparing with the fit to $xF_3$ data only,
the HT parameters errors are  decreasing.
Within the errors, the parameters, which describe   
the boundary distributions, are compatible 
with the outcomes of the similar fits from Ref.\cite{SELIG}.
The coefficients of $H_3(x)$ are in
agreement with the NLO results of Ref.\cite{KKPS}
and the behaviour of $H_2(x)$ qualitatively reproduce the 
HT contribution to $F_2$,  obtained 
from the combined fits of the SLAC/BCDMS data on $F_2$, performed 
in Refs.\cite{VM,Aln}.

When the matching scale $M_5$ was changed from $m_b$ to $6.5m_b$,
the value of $\alpha_s(M_Z)$ shifted down by 0.0052 and  then we ascribe to 
$\alpha_s(M_Z)$ the theoretical error of 0.0026 due to uncertainty 
of b-quark threshold matching. 
This uncertainty is in agreement with the results of the 
NLO Jacobi-polynomial fits of the CCFR data obtained within
so-called spline $\overline{MS}$ prescription \cite{ShSM}.
One more source of the theoretical uncertainty, which is
due to the truncation of higher QCD orders,  was evaluated
following the way, proposed in Ref. \cite{VM}. 
In accordance with their procedure
one can  introduce renormalization scale $k_R$
into QCD evolution equations in the way, illustrated on the example of 
NS evolution:
\begin{displaymath}
\frac{dxq^{NS}}{d\ln Q}=\frac{\alpha_s(k_RQ)}{\pi}\int^1_x
dz\biggl\{P^{NS,(0)}_{qq}(z)+
\end{displaymath}
\begin{equation}
+\frac{\alpha_s(k_RQ)}{2\pi}
\Bigl[P^{NS,(1)}_{qq}(z)+\beta_0P^{NS,(0)}_{qq}(z)\ln(k_R)\Bigr]
\biggr\}\frac{x}{z}q^{NS}(x/z,Q),
\end{equation}
where $P^{NS,(0)}$ and $P^{NS,(1)}$ denote the  LO and the NLO parts of 
the splitting function $P^{NS}$.
The dependence of the results on $k_R$ would signal an incomplete 
account of 
the perturbation theory effects.
The shift of $\alpha_s(M_Z)$ 
resulting from the reasonable variation of $k_R$ leads to 
an additional error of over 0.003 due to the renormalization scale
uncertainty. Taking $Q_0^2=20~GeV^2$ as the initial scale,  we have
checked that our results are rather stable to the variation 
of the factorization point.   

The NLO value we fare presenting as the main result is thus
\begin{equation}
\alpha_s(M_Z)=0.1222\pm 0.0048 (stat+syst) \pm 0.0040 (thresh+ren.scale) 
\end{equation}
It  differs a bit from the NLO  value 
$\alpha_s(M_Z)=0.119 \pm 0.002 (stat+syst) \pm 0.004(theory)$,
obtained in the CCFR
analysis \cite{CCFR}. The increase of the
experimental errors is due to the fact that while CCFR group used 
model-dependent form of the HT contributions, we are considering them as
the additional free parameters and are extracting them from the fits.  

In order to try to  decrease further the errors,  we repeated the
fits of the combined $xF_3$ and $F_2$ data, using the less stringent cut
$Q^2>1~GeV^2$. 
The obtained results are presented in the third column of Table 2.
In these fits the parameters $b_s$ and $b_G$ were
released since their values turned out to be statistically different 
from zero. We found, that the values of $\alpha_s$ and $b_G$ are 
correlated (the correlation coefficient is equal to --0.65). When we fixed 
$b_G=0$, the value $\alpha_s(M_Z)=0.1172 \pm 0.0029$ was obtained, and when 
we kept $b_G$ as the free parameter, we
obtained low value of $\alpha_s(M_Z)=0.1131\pm0.0045$.
The analogous effect of correlations was observed for the fit with the cut
$Q^2>5~GeV^2$, although with less statistical significance.
It should be underlined, that when we released 
$b_G$ in the fit with the cut $Q^2>1~ GeV^2$ we faced another problem: 
its value turned out to be 
\begin{equation}
b_G=0.092\pm 0.073,
\end{equation}
which is in the evident contradiction with the results, obtained 
in the analysis of HERA data (for example the combined analysis of DIS
from HERA and CERN-SPS data results in 
the value $b_G=-0.267\pm 0.043$  
\cite{ALEK}, while in the framework of MRST parametrization the value  
$b_G=-1.08$\cite{MRST} was obtained).This problem
might be related to the well-known discrepancy
between CCFR and NMC/BCDMS data at small $x$. 
  
{\bf Conclusion}

In conclusion, we would like to stress that in order to perform 
similar analysis at the NNLO level it is necessary to calculate the 
yet unknown Altarelli-Parisi kernels to the corresponding 
DGLAP equations. Therefore, we are unable to obtain the results, 
similar to the NNLO  ones of Refs.\cite{KKPS,KPS}. We hope that future 
progress of theoretical calculations  will allow us to generalize our results 
to the NNLO approximation.

{\bf Acknowledgements}

We are grateful to W. Bernreuther, G. Parente,
G. Ridolfi and A.V. Sidorov for the interest in this our work and 
for discussions.
Our work was partly supported by the Russian Fund for Fundamental
Research, Grant N 96-02-18897. The work of the second author  was also
supported by RFFI Grant N 96-01-01860.  

This work was completed during the stay of one of us (ALK) in the Theory 
Division of CERN. The warm hospitality   and ideological
support of its members should be gratefully acknowledged. 

The work on the final version of the paper was done within the 
scientific program of the Project N99-02-16142, submitted to the 
Russian Foundation of Fundamental Research.


\begin{thebibliography}{99}

\bibitem{BB1}
E.L. Berger and S.J. Brodsky, {\it Phys. Rev. Lett.} {\bf 42} (1979)
940;\\
J.F. Gunion, P. Nason and R. Blankenbecler, {\it Phys. Rev.} {\bf D29}
(1984) 2491.

\bibitem{AB}
L. F. Abbott and R.M. Barnett, {\it Ann. Phys.} {\bf 129} (1980) 276;\\
L.F. Abbott, W.B. Atwood and R.M. Blankebecler, {\it Phys. Rev.} 
(1980) 582.

\bibitem{Isaev}
V.A. Bednyakov, P.S. Isaev and S.G. Kovalenko, {\it Yad. Fiz} {\bf 40} 
(1984) 770; ({\it Sov. J. Nucl. Phys.} {\bf 40} (1984) 494).

\bibitem{Buras}
A.J. Buras, {\it Rev. Mod. Phys.} {\it 52} (1980) 199.

\bibitem{BB}
M. Beneke and V.M. Braun, {\it Phys. Lett.} {\bf B348} (1995) 513.

\bibitem{DMW}
Yu. L.Dokshitzer, G. Marchesini and B.R. Webber, {\it Nucl. Phys.} 
{\bf B469} (1996) 93.

\bibitem{DW}
M. Dasgupta and B.R.~Webber, {\it Phys. Lett.} {\bf B382} (1996) 273.

\bibitem{Stein}
M. Mauel, E. Stein, A. Sch\"{a}fer and L. Mankiewicz, 
{\it Phys. Lett.} {\bf B401} (1997) 100.


\bibitem{AZ}
R. Akhoury and V.I. Zakharov, Proc. of QCD-96 Int. Workshop, 
Montpellier, ed. S. Narison, {\it Nucl. Phys. Proc. Suppl.} 
{\bf 54A} (1997) 217; (hep-ph/9610492)

\bibitem{BBM}
M. Beneke, V.M. Braun and L. Magnea, {\it Nucl. Phys.} {\bf B497} (1997)
297.

\bibitem{B}
V.M. Braun, Proc. XXX Recontre de Moriond "QCD and High Energy Hadronic
Interactions", ed. J. Tran Than Van, Editions Frontieres (1995) p.271
(hep-ph/9505317);\\
V.I. Zakharov, Talk at QCD-98 Worksop, Montpellier, 1998
(hep-ph/9811294).

\bibitem{Beneke}
M. Beneke, preprint CERN-TH-98-233 (hep-ph/9807443).


\bibitem{CCFR} 
CCFR collaboration, W.G. Seligman et al., 
               {\it  Phys. Rev. Lett.} {\bf 79} (1997) 1213.

\bibitem{SELIG} W.G. Seligman, Thesis, Report No. Nevis-292, 1997.

\bibitem{KKPS} A.L. Kataev, A.V. Kotikov, G. Parente and  A.V. Sidorov,
{\it Phys. Lett.} {\bf 417B} (1998) 374.


\bibitem{VZ}
W.L. van Neerven and E.B. Zijlstra, {\it Nucl. Phys.} {\bf B383} (1992)
235.

\bibitem{LRV}
S.A. Larin, T. van Ritbegen and J.A.M. Vermaseren, 
{\it Nucl. Phys.} {\bf B427} (1994) 41;\\
S.A. Larin, P. Nogueira, T. van Ritbergen and J.A.M. Vermaseren, 
{\it Nucl. Phys.}{\bf B492} (1997) 338. 


\bibitem{PKK}
G. Parente, A.V. Kotikov and V.G. Krivokhizhin, {\it Phys. Lett.} 
{\bf B333} (1994) 190. 

\bibitem{KKPS1}
A.L. Kataev, A.V. Kotikov, G. Parente and A.V. Sidorov, {\it Phys. Lett.}
{\bf B388} (1996) 179.

\bibitem{VM}
M. Virchaux and A. Milsztain, {\it Phys. Lett.} {\bf B274}(1992) 221.

\bibitem{KKPS2}
A.L.Kataev, A.V. Kotikov, G. Parente and A.V. Sidorov, 
Proc. QCD-97 Workshop, Montpellier, July 1997, ed. S. Narison,
{\it Nucl. Phys. Proc. Suppl.} {\bf 64} (1998) (hep-ph/9709509)

\bibitem {KPS}
A.L. Kataev, G. Parente and A.V. Sidorov, hep-ph/9809500.


\bibitem{ALEK}
S.I. Alekhin, Report No. IHEP 96-79, hep-ph/9611213; to 
appear in {\it Eur. Phys. J.}


\bibitem{Jacobi1}
G. Parisi and N. Sourlas, {\it Nucl. Phys.} {\bf B151} (1979) 421.

\bibitem{Jacobi2}
J. Ch\'yla and J. Ramez, {\it Z. Phys.} {\bf C31} (1986) 151.

\bibitem{Jacobi3}
V.G. Krivokhizin et al., {\it Z. Phys.} {\bf C36} (1987) 51; ibid.
{\bf C48} (1990) 347.

\bibitem{DGLAP}
V.N. Gribov and L.N. Lipatov, {\it Sov. J. Nucl. Phys.} 
{\bf 15} (1972) 438;\\
L.N. Lipatov, {\it Sov. J. Nucl. Phys.} {\bf 20} (1975) 94;\\
G. Altarelli and G. Parisi, {\it Nucl. Phys.} {\bf 126} (1977) 298;\\
Yu. L. Dokshitzer, {\it JETP} (1977) 641.


\bibitem{BENCH}
 J. Bl\"{u}mlein  et al., 
    Report No. DESY 96-199, hep-ph/9609400;
Proc. "Hamburg 1995/1996, Future physics at HERA", 1996, p. 23.

\bibitem{Aln}
S.I. Alekhin, Report No. IHEP 98-67, hep-ph/9809544.

\bibitem{GGK}
E.~Gardi, G. Grunberg and M. Karliner, hep-ph/9806462;\\
B.A. Magradze, Talk at the Int. Conf. "Quarks-98", Suzdal, May 1998,
hep-ph/9808247.

\bibitem{BW}
W. Bernreuther and W. Wetzel, {\it Nucl. Phys.} {\bf B197} (1982) 228;
ibid. {\bf B513} (1998) 758 (Err.);\\
S.A. Larin, T. van Ritbegen and J.A.M. Vermaseren, 
{\it Nucl. Phys.} {\bf B428} (1995) 278;\\
K.G. Chetyrkin, B.A. Kniehl and M. Steinhauser, {\it Phys. Rev. Lett.}
{\bf 79} (1997) 2184.


\bibitem{BN}
J. Bl\"{u}mlein and W.L. van Neerven, preprint DESY 98-176 
(hep-ph/9811351).


\bibitem{DO}
A. Devoto, D.W. Duke, J.F. Owens and R.G. Roberts,
{\it Phys. Rev.} {\bf D27} (1983) 508.



\bibitem{MS}
G. Curci, W. Furmanski and R. Petronzio, 
    {\it Nucl. Phys.} {\bf B175} (1980) 27;\\
W. Furmanski and R. Petronzio, {\it Z. Phys.} {\bf C11} (1982) 293.
 

\bibitem{GLS} D.J. Gross and  C.H. Llewllyn Smith, 
{\it Nucl. Phys.} {\bf B14} (1969) 337.

\bibitem{TMC}
H. Georgi and H.D. Politzer, {\it Phys. Rev.} {\bf D14} (1976) 1829.


\bibitem{ShSM}
D.V. Shirkov, A.V. Sidorov and S.V. Mikhailov, hep-ph/9707514

\bibitem{MRST}
A.D. Martin, R.G. Roberts, W.J. Stirling and R.S. Thorne,
{\it Eur.Phys.J.} {\bf C4} (1998) 463.

\end{thebibliography}
\end{document}